\documentclass[9pt,shortpaper,twoside,web]{ieeecolor2}
\usepackage{generic}
\usepackage{cite}
\usepackage{amsmath,amssymb,amsfonts}
\usepackage{graphicx}
\usepackage{textcomp}

\usepackage{algorithm}
\usepackage[outdir=./]{epstopdf}

\usepackage[utf8]{inputenc}
\usepackage{authblk}
\usepackage[noend]{algpseudocode}
\usepackage{mathtools,amsmath,amsfonts,amssymb, mathrsfs}
\usepackage{booktabs, makecell, multirow}
\usepackage{soul}
\usepackage{etoolbox}
\usepackage{comment}
\usepackage[table, dvipsnames]{xcolor}
\DeclareRobustCommand{\legendsquare}[1]{%
  \textcolor{#1}{\rule{2ex}{2ex}}%
}

\usepackage[switch,columnwise]{lineno}

\usepackage{pifont}
\usepackage[acronym]{glossaries}
\newacronym{ecog}{ECoG}{electrocorticography}
\newacronym{bci}{BCI}{brain-computer interface}
\newacronym{BMI}{BMI}{brain-machine interface}
\newacronym{AI}{AI}{artificial intelligence}
\newacronym{rnn}{RNN}{recurrent neural network}
\newacronym{eeg}{EEG}{Electroencephalography}
\newacronym{snr}{SNR}{signal-to-noise ratio}
\newacronym{lstm}{LSTM}{long short-term memory}
\newacronym{svm}{SVM}{support vector machine}
\newacronym{rfc}{RFC}{random forest classifier}
\newacronym{emg}{EMG}{electromyogram}
\newacronym{psd}{PSD}{power spectral density}
\newacronym{bop}{Bop}{binary optimiser}
\newacronym{ers}{ERS}{event related synchronisation}
\newacronym{std}{STD}{standard deviation}
\newacronym{lda}{LDA}{linear discriminant analysis}
\newacronym{lightgbm}{LightGBM}{light gradient boosting machine}
\newacronym{dbs}{DBS}{Deep Brain Stimulation}
\newacronym{nam}{NAM}{Neural Additive Model}
\newacronym{osd}{OSD}{Optimal Seizure Detection}
\newacronym{ttca}{TTCA}{Test-time classifier adjustment}
\newacronym{tta}{TTA}{Test-Time Adaptation}
\newacronym{sgd}{SGD}{stochastic gradient descent}
\newacronym {T3A}{T3A}{Test-Time Templates Adjuster}
\newacronym {chb}{CHB-MIT}{Children’s Hospital Boston and the Massachusetts Institute
of Technology}
\newacronym {ML}{ML}{Machine Learning}
\newacronym {SVM}{SVM}{ Support Vector Machines }
\newacronym {DNN}{DNN}{ Dense Neural Network}
\usepackage{etoolbox}
\usepackage{lettrine}
\usepackage{epstopdf}
\usepackage{orcidlink}

\usepackage{xcolor}
\usepackage{hyperref}

\hypersetup{
    colorlinks=true,
    linkcolor=violet, 
    citecolor=blue,
     urlcolor   = black
}

\makeatletter
\renewcommand\citepunct{,\penalty\@m\hskip.13em\relax}
\renewcommand\@cite[2]{\textcolor{blue}{[{#1\if@tempswa , #2\fi}]}}
\makeatother

\def\BibTeX{{\rm B\kern-.05em{\sc i\kern-.025em b}\kern-.08em
    T\kern-.1667em\lower.7ex\hbox{E}\kern-.125emX}}
\begin{document}
\title{MT-NAM: An Efficient and Adaptive Model for Epileptic Seizure Detection}

\author{Arshia Afzal \orcidlink{0000-0001-9927-2492}, \and Volkan Cevher \orcidlink{0000-0002-5004-201X}, \IEEEmembership{Fellow,~IEEE} and Mahsa Shoaran \orcidlink{0000-0002-6426-4799}, \IEEEmembership{Senior Member,~IEEE}
\thanks{ Arshia Afzal is with the Integrated Neurotechnologies Laboratory Laboratory and Information and Inference Systems, Institutes of Electrical and Micro Engineering and Neuro-X, EPFL, Switzerland. e-mail: (arshia.afzal@epfl.ch)}
\thanks{ Mahsa Shoaran is with the Integrated Neurotechnologies Laboratory, Institutes of Electrical and Micro Engineering and Neuro-X, EPFL, 1202 Geneva, Switzerland.  e-mail: (mahsa.shoaran@epfl.ch)}
\thanks{ Volkan Cevher is with the Laboratory for Information and Inference Systems, EPFL, 1015 Lausanne, Switzerland. e-mail: (volkan.cevher@epfl.ch)}
}

\maketitle

\begin{abstract}
Enhancing the accuracy and efficiency of  machine learning algorithms employed in  neural interface systems is crucial for advancing next-generation intelligent therapeutic devices.  
However, current systems often utilize basic machine learning models that do not fully exploit the natural structure of brain signals. Additionally, existing learning models used for neural signal processing often demonstrate low speed and efficiency during  inference. To address these challenges, this study introduces Micro Tree-based NAM (MT-NAM), a distilled model based on the recently proposed Neural Additive Models (NAM). The MT-NAM achieves a remarkable 100$\times$ improvement in inference speed compared to standard NAM, without compromising accuracy. 
We evaluate our approach on the CHB-MIT scalp EEG dataset, which includes recordings from 24 patients with varying numbers of sessions and seizures. NAM achieves an 85.3\% window-based sensitivity and 95\% specificity. Interestingly, our proposed MT-NAM  shows only a 2\% reduction in sensitivity compared to  the original NAM. 
To regain this sensitivity, we utilize a test-time template adjuster (T3A) as an update mechanism, enabling our model to achieve higher sensitivity during test time by accommodating transient shifts in neural signals. With this online update approach, MT-NAM achieves the same sensitivity as the standard NAM  while achieving approximately 50$\times$ acceleration in inference speed.
\end{abstract}

\begin{IEEEkeywords}
Seizure Detection, Neural Additive Model (NAM), Machine Learning, Accurate, Inference, Test-Time Adaptation, Decision Tree, \gls{eeg}, Knowledge Distillation, Sensitivity. 
\end{IEEEkeywords}

\section{Introduction}
\label{sec:introduction}
\lettrine{B}{rain} disorders, including epilepsy, pose significant challenges worldwide, necessitating the exploration of innovative approaches for diagnosis and treatment. Seizures are among the most common neurological emergencies globally \cite{strein2019prevention,tang2021self,li2022graph,ho2023self}, and can become chronic, as  in the case of epilepsy—a neurological disorder affecting over 50 million  worldwide~\cite{WHO}. Seizure detection from EEG has emerged as a promising technique to identify abnormal brain activity associated with seizures. However, traditional machine learning (ML) models used for this task face computational and memory constraints, limiting their applicability in real-time scenarios \cite{asif2020seizurenet,bajpai2021automated}. 

Seizures are commonly characterized by their focal or non-focal nature \cite{focal}, with activity either starting in a single channel and spreading or remaining confined to specific channels.  In contrast to existing models that aggregate data from all channels for predictions \cite{asif2020seizurenet,soul,resot,tang2021self}, this method may be less effective for seizure detection because it fails to account for the localized and often channel-specific nature of seizure events. Moreover, existing ML-embedded Systems-on-Chip (SoCs) for wearable or implantable applications predominantly rely on efficient ML architectures such as ridge regression \cite{circ2}, SVM-based models \cite{svm,circ1}, decision trees \cite{2022JSSC_Uisub,shoaran2018energy, shoaran2016hardware}, and logistic regression (LR) \cite{soul}. While these designs excel in terms of power consumption and area efficiency, they often fall short in terms of accuracy when compared to deep neural structures used for seizure detection \cite{asif2020seizurenet}. 
Additionally, newer and more innovative network architectures have recently been introduced  \cite{attention}, which could potentially be suitable for seizure detection tasks if they are optimized for hardware efficiency \cite{taghavi2019hardware,afzal}. 
We hypothesize that training a deep model on seizure data and then distilling this knowledge \cite{knowladgedis} into a smaller model can simultaneously improve both accuracy and efficiency.

Furthermore, it is important to note that seizure dynamics can evolve over time, leading to reduced performance in machine learning models. Recent work  \cite{soul} has addressed this by applying test-time updates to a simple LR model, improving sensitivity to ictal samples. Additionally, there is potential in exploring more advanced update rules, successfully applied in other domains \cite{tent, etta} to enhance model performance in the context of epileptic seizure detection.

This work addresses the key challenges involved in developing hardware-efficient models for seizure detection, where balancing accuracy and resource constraints is crucial. First, we introduce a Neural Additive Model (NAM) \cite{nam} tailored for seizure detection, which effectively captures the non-linear relationships in the data. To improve computational efficiency, we propose a distilled version, the Micro Tree-based NAM, which reduces parameters and speeds up processing without sacrificing performance. Lastly, we enhance the model's adaptability through the T3A update mechanism, allowing for dynamic adjustments during test-time. 
The key contributions of this work are as follows:
\begin{itemize}
    \item We employ the Neural Additive Model (NAM) as a novel approach for seizure detection. NAM is capable of learning individual feature functions for each input feature, while the final prediction is made by summing these outputs, which aligns well with the diverse and distributed nature of seizure events. Our method  achieves an window-based sensitivity of 85\%, an event-based sensitivity of 100\%, and a specificity of 96\% for seizure detection across all patients on  the CHB-MIT  dataset, outperforming the previous benchmarks for computationally efficient seizure detection \cite{2022JSSC_Uisub,soul,cnn1}.
    To the best of our knowledge, this is the first application of a Neural Additive Model in seizure detection.

    \item We further present Micro Tree-based NAM (MT-NAM), a model developed through knowledge distillation from NAM, where the feature functions are replaced with regression trees. This approach achieves a 100$\times$ improvement in inference speed compared to NAM and significantly reduces the model's computational demands. 

    \item We utilize the T3A \cite{ttca} method to adapt the model during test-time, offering more stable results compared to alternative approaches. While MT-NAM effectively reduces computational complexity and accelerates inference, it leads to a slight decrease in sensitivity to seizures, which is critical since missing ictal samples is unacceptable. To address this, we develop a more memory-efficient version of the T3A \cite{ttca} approach tailored for binary classification tasks. This enhanced model strikes an attractive balance,  maintaining the sensitivity  of the original NAM while achieving a 50$\times$ speed increase during test-time. 
    
\end{itemize}

\begin{figure*}
\begin{minipage}[t]{2\columnwidth} 
 \includegraphics[width=\textwidth]{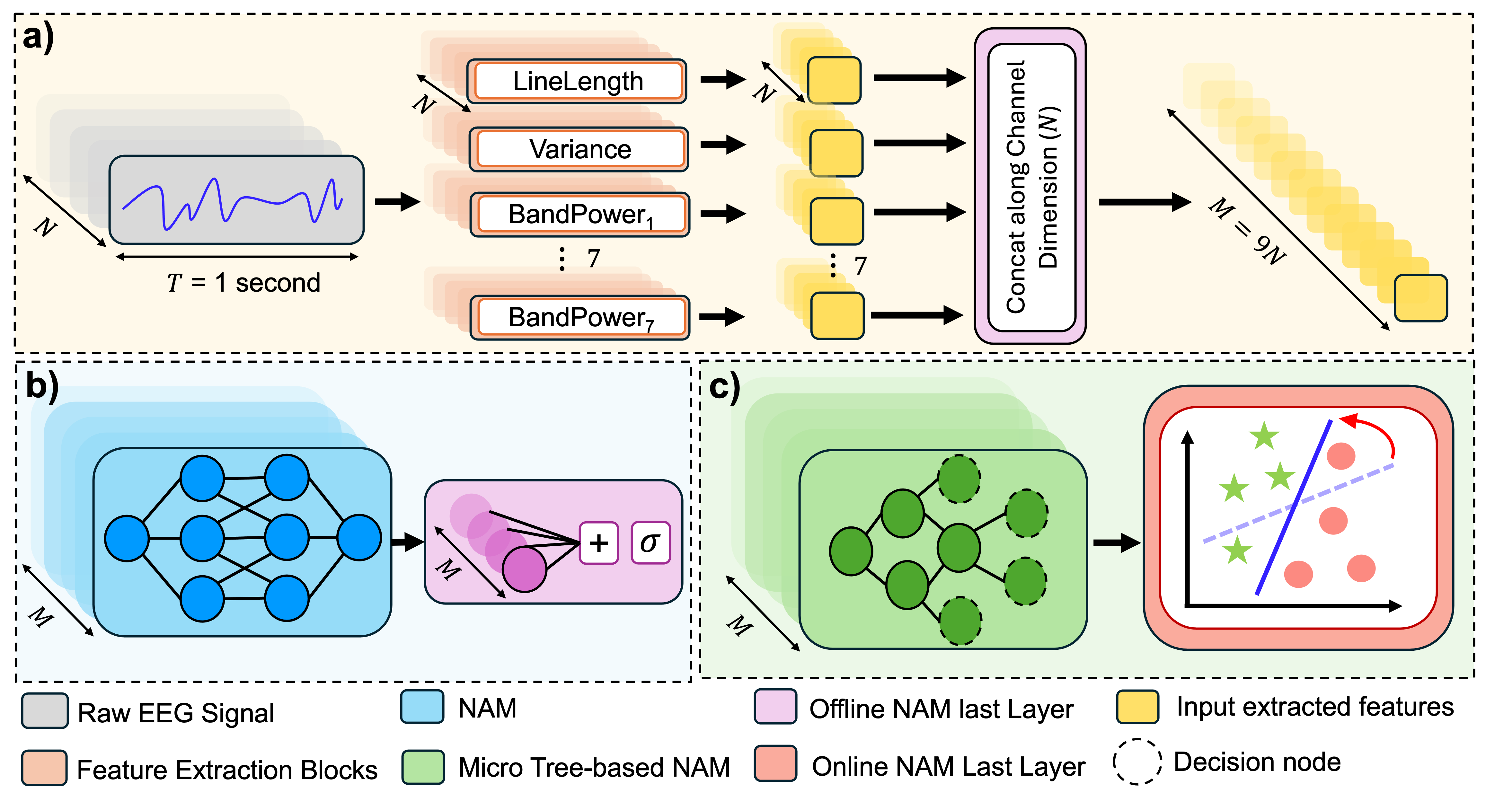}
  \caption{The proposed model framework. \textbf{(a)} \textit{Feature extraction.} During this phase 7 band power features aligned with Linelenght and Variance is extracted from each channel of raw EEG signal for every 1 second window size resulting in 9 features per channel and totally $M=9N$ number of features with $N$ being the channel number. \textbf{(b)} \textit{Training NAM.} During training phase NAM is trained using the extracted features and offline prediction is applied for seizure detection. \textbf{(c)} \textit{MT-NAM Inference} During inference the distilled version of NAM namely, Micro Tree-based NAM (MT-NAM) and online T3A update is used for predicting the seizures. \textbf{Color Codes:} \legendsquare{orange!10} Indicates the feature extraction paradigm (\textbf{a}), \legendsquare{cyan!10} is the training paradigm (\textbf{b}) and \legendsquare{green!10} is the inference paradigm during testing (\textbf{c}). }
  \label{fig:fig1}
\end{minipage}
\end{figure*}


The remainder of this paper is organized as follows. Section \ref{sec:2} provides a detailed explanation of the methodology used for architecture design, knowledge distillation, and the online update process, while Section \ref{sec:3} covers the experimental setup and presents the results of the experiments. Section \ref{sec:relwork} describes the related work on seizure detection and section \ref{conc} is the conclusion of our paper.

\section{Methods}
\label{sec:2}
\subsection{Seizure detection problem formulation}
The primary objective of  seizure detection  is to determine whether a seizure is present within a given  EEG window. Here, we focused on evaluating our model's performance with  1-second windows to address more realistic scenarios. This window is short enough to allow timely intervention through closed-loop neurostimulation \cite{2022JSSC_Uisub, circ2, shoaran201616, zhu2021closed, shoaran2024intelligent
}, potentially suppressing impending seizures, while also being long enough to capture low-frequency content crucial for accurate detection \cite{window}.

\subsection{Neural additive models for seizure detection} 
In this work we employ a \gls{nam} \cite{nam} for seizure detection, as depicted in {Fig.~\ref{fig:fig1}}. To this end, we extract the features of line length, variance, and spectral band power from each \gls{eeg} channel \cite{OSD}, which we refer to as input features (distinct from the feature functions used in \gls{nam}). 

NAM learns an individual function for each input feature and combines these learned functions in the final layer to make predictions. This design of the model incorporates two key advantages: it provides interpretability by offering clear insights into the learned feature functions for each input, as demonstrated in the context of seizure detection where these functions are visualized in Figure \ref{fig:featfunc}. Additionally, by treating each learned feature function independently, this approach enables the exploration of novel distillation methods to enhance model efficiency, as each feature function can be regarded as a black-box learned function, opening new avenues for optimization.

Each input feature is fed into a neural network called the feature network, and the final prediction is obtained by summing the outputs of these feature networks. The prediction is formulated as follows:
\begin{equation} \label{eq:1}
\hat{y}_t = \sigma \left( \sum_{c=1}^{N} \sum_{i=1}^{F} f_{\theta_{i,c}}(x^{i,c}_{t}) \right ) ,
\end{equation}
where $\hat{y}_t$ represents the NAM prediction at time window $t$, $x^{i,c}_{t}$ is the $i$'th feature extracted from channel number $c$ with $c \in [1,...,N]$ and $i \in [1,...,F]$ , $ f_{\theta_{i,c}}(.)$ is the function learned by \gls{nam} feature network for $x^{i,c}_{t}$ input, $\sigma$ denote the sigmoid function, $N$ denote the total number of channels (for this work 23) and $F$ denotes the total number of extracted features per-channel (for this work 9). By concatenating the input features $x^{i,c}_{t}$ and the outputs of the \gls{nam} feature function $ f_{\theta_{i,c}}(x^{i,c}_{t})$ into $M$-dimensional vectors $ f_\theta(x_t) \in R^{1 \times M}$ and $x_t \in R^{1 \times M}$ with $M =N \times F$, the  equation (\ref{eq:1}) can be expressed in a simplified form as follows:
\begin{equation} \label{eq:2}
\hat{y}_t= \sigma (f_\theta(x_t) \cdot  \mathbf{1} ) ,
\end{equation}
where $\mathbf{1}, = [1,1,...1] \in R^{1 \times M}$ and $\cdot$ represents the inner product.

\subsection{Tree-Based Knowledge Distillation}
Knowledge distillation is a technique in which a deep, pre-trained model is used to train a smaller (or otherwise optimized) model that achieves comparable accuracy with greater efficiency \cite{knowladgedis}. Typically, this process follows a teacher-student framework, where the smaller model learns not directly from the label distribution but from the output distribution of the teacher (deeper) model, a process known as \textit{teacher-student distillation} \cite{ts}.

In this work, we introduce a novel type of distillation, where, to the best of our knowledge, we are the first to {distill the knowledge of a neural network into a decision tree by approximating the feature functions} of a NAM. We refer to this technique as \textit{tree-based distillation}. Unlike traditional methods that rely on backpropagation, this approach uses a decision tree as the student model to approximate the teacher’s feature functions, rather than learning its output distribution. This enables the tree model to be obtained in a deterministic manner.

\subsection{Micro Tree-based NAM}
In the original \gls{nam} structure \cite{nam}, each input feature is associated with a learned function, and the final prediction is obtained by summing the outputs of these functions and applying a sigmoid activation. However, the computational and memory requirements for  calculating the outputs of each feature network can be substantial, leading to increased inference latency. To address these challenges, {we propose \textit{tree-based distillation}, an efficient approximation technique for the learned feature functions by employing micro decision trees, which we refer to as Micro Tree-based NAM (MT-NAM).}

By leveraging the interpretability of \gls{nam}, we replace each feature function with a small regression tree, limiting its depth to a maximum value of $d$. This approach significantly reduces the computational and memory demands during inference while maintaining competitive performance. The algorithm for approximating  NAM feature functions using micro decision trees is outlined in Algorithm \ref{alg:mtnam}.

\begin{algorithm}

\caption{Distilled Micro Tree-based NAM (MT-NAM)}\label{euclid}
\textbf{Input:} \gls{nam} feature functions, $f_{\theta}$,  and the training set $\mathcal{T}$ with $N$ samples.
\\
\textbf{Output:} NAM's approximated feature functions $\tilde{f_{\theta}}$.
\begin{algorithmic}
\State $\tilde{f_{\theta}} \gets \{\}$
\For {$ f_{\theta_{j}}$ \textbf{in} $f_{\theta}$}
\State $\tilde{f_{\theta_j}} \gets$  Regression Decision Tree (max-depth = $d$).
\State  \{$x_j^1,...,x_j^{M} \} \gets$   Sample $j$'th input feature of data points in $\mathcal{T}$.
\State $\mathbf{S}_j \gets \bigl\{\big(z , f_{\theta_j}(z)\big) : z \in \{x_j^1,...,x_j^{M}\}\bigl\}$   .
\State Fit $\tilde{f}_{\theta_j}$ to $\mathbf{S}_j$.
\State $\tilde{f_{\theta}}$.append ($\tilde{f_{\theta_j}}$)

\EndFor
\State \textbf{return} $\tilde{f_{\theta}}$
\end{algorithmic}
\label{alg:mtnam}
\end{algorithm}

This way each feature network of NAM is replaced by small regression decision tree (\ref{fig:fig1}) which is light weighted and memory efficient, and by increasing the depth of the decision tree the final approximation will become closer to actual NAM feature function which controls the level of approximation for \textit{tree-based distillation}.

\subsection{Adaptive NAM and MT-NAM}
In general,  a model's adaptability to unseen domains during  the testing phase is achieved through different techniques known as \gls{tta}\cite{ttt,testtimeadaptationsurv}. In practice, \gls{tta}  offers two primary methods for adapting the classifier during test time, which include:

\textbf{1) Back propagation-based model update.} These methods modify  model weights during test-time using stochastic gradient descent (SGD): 
\begin{equation}\label{eq:3}
    \theta_{t+1} = \theta_t - \eta \nabla \mathcal{L}(x_t,\theta_t) ,
\end{equation}
Here, $\theta_{t+1}$ represents the updated weight, $\theta_t$ denotes the current weight of the model, $\eta$ signifies the learning rate, and $x_t$ represents the input at time point $t$. The goal is to minimize the loss function $\mathcal{L}(x_t,\theta_t)$ during test-time, which is typically  an unsupervised loss (as an example entropy \cite{tent}) due to the absence of labels during the test phase. 

Notably, \cite{ttt} and \cite{ttt++} serve as examples of back propagation-based methods that train a source model using both supervised and self-supervised objectives, and later adapt the model during test time using a self-supervised objective. However, this approach relies on specific assumptions about the training process, which may not always be feasible in practice. As a result, fully test-time adaptation models like \cite{tent,Robust} focus on adapting the model exclusively using test data during inference. 
 
While domain generalization can be achieved by optimizing model parameters during test time, it is important to address potential challenges introduced by stochastic optimization methods. These include catastrophic forgetting \cite{catasforget}, where the model loses previously learned information, and overfitting to specific data classes due to sample redundancy during test time. Although various studies \cite{catasforget, etta} have proposed solutions to mitigate these issues, we found that applying these methods to update our model resulted in suboptimal performance during test time

\textbf{2) Back propagation-free model update.} Unlike back propagation-based update techniques, back propagation-free methods do not rely on gradient-based optimization for model updates \cite{freeparam,ttca}. Instead, they utilize optimization-free strategies to adjust the output of the last layer in the model, allowing adaptation to unseen domains during test-time:
\begin{equation} \label{eq:4}
    y_t = g(f_\theta(x_t) , x_t) ,
\end{equation}
where ${y_t}$ represents the prediction of the adjusted classifier at time $t$, $g(.)$ denotes the adjustment function, $f_\theta(.)$ is the output of the classifier with pre-trained $\theta$ parameters, and $x_t$ refers to the input at time point $t$.

As depicted in (\ref{eq:4}), the model parameters ($\theta$) remain fixed during  test time, with adjustments made only to the model's last layer. While the approach in \cite{freeparam} requires access to all previous input samples during test time to update each prediction, \cite{ttca} depends only on the mean of the previous samples, making it more efficient for real-time applications such as seizure detection.

\textbf{Online adaptation of NAM and MT-NAM.} In this study, we employed a simplified version of the \gls{T3A} method to update the last layer of our model. We selected T3A for its simplicity in handling binary classification (0 for non-ictal and 1 for ictal labels) \cite{ttca}. Its back propagation-free nature makes it particularly suitable for tree-based models such as MT-NAM. After training our model on the source domain, we applied the \gls{T3A} algorithm to update the support sets $\mathbb{S}^0_{t-1}$ and $\mathbb{S}^1_{t-1}$ as follows:
\begin{equation} \label{eq:5}
(\mathbb{S}^0_{t} , \mathbb{S}^1_{t}) = \begin{cases*}
  (\mathbb{S}^0_{t-1} \cup \frac{f_\theta(x_t)}{||f_\theta(x_t)||_2} , \mathbb{S}^1_{t-1}) & if $\hat{y}_t$ = 0,\\          
  (\mathbb{S}^0_{t-1},   \mathbb{S}^1_{t-1} \cup  \frac{f_\theta(x_t)}{||f_\theta(x_t)||_2}) & if $\hat{y}_t$ = 1,
\end{cases*}
\end{equation}
\\
In the above equation, the support sets $(\mathbb{S}^0_{0} , \mathbb{S}^1_{0})$ are initialized as $ \bigl( \bigl\{\frac{\mathbf{1}}{\sqrt{N}}  \bigl\} ,\bigl\{\frac{-\mathbf{1}}{\sqrt{N}}  \bigl\} \bigl)$ based on the summation of all feature functions in the last layer of \gls{nam} (eq. \ref{eq:2}), where $\mathbf{1}$ denotes an all-ones vector. 
\\
\\
The prediction adjustment is computed using:
\begin{equation} \label{eq:6}
y_t = \frac {\exp \left(f_\theta(x_t).\mu^1_{t}\right)}{ \smashoperator \sum _{k=0,1} {\exp \left(f_\theta(x_t).\mu^k_{t}\right) }} = \sigma \left( f_\theta(x_t) \cdot (\mu^1_{t} - \mu^0_{t}) \right),
\end{equation}
\\
where $\mu^0_{t} , \mu^1_{t}$ represent the centroids of $\mathbb{S}^0_{t}$ and $ \mathbb{S}^1_{t}$ at time $t$:
\begin{equation} \label{eq:7}
\mu^0_{t} = \frac{1}{|\mathbb{S}^0_{t} |} \smashoperator \sum_{z \in \mathbb{S}^0_{t} } {z} \hspace{0.2cm }\text{and} \hspace{0.2cm } \mu^1_{t} = \frac{1}{|\mathbb{S}^1_{t} |} \smashoperator \sum_{z \in \mathbb{S}^1_{t} } {z},
\end{equation}

By defining $n^1_t = |\mathbb{S}^1_{t}|$ and $n^0_t = |\mathbb{S}^0_{t}|$, we can adjust the final predictions using the support set parameters $\mu^1_{t-1}$, $\mu^0_{t-1}$, $n^1_{t-1}$, and $n^0_{t-1}$. This approach eliminates the need to store the $(\mathbb{S}^0_{0}, \mathbb{S}^1_{0})$ sets, making the algorithm more memory-efficient (details described in Algorithm \ref{alg:onlinenam}).

To avoid updating with non-confident samples, we measure the prediction entropy of the \gls{nam}, $H(\hat{y}_t) = -\big( \hat{y}_t\log (\hat{y}_t) + (1-\hat{y}_t)\log (1-\hat{y}_t)\big)$ to filter out unreliable labeled data with entropy exceeding a threshold (i.e., $H_0$). The value of $H_0$ is tuned during the validation process.

\begin{algorithm}
\caption{Adaptive NAM for Seizure Detection}\label{euclidd}
\textbf{Input:} \gls{nam}/MT-NAM feature functions in a vectoized form $f_\theta$, input sample at time point $t$ in a vectoized form ($x_t$), support set parameters available at this time ($\mu^1_{t-1} , \mu^0_{t-1} , n^1_{t-1} , n^0_{t-1}$).
\\
\textbf{Output:} Adjusted prediction for input $x_t$
\begin{algorithmic}
\State $\hat{y} = \sigma(f_\theta(x_t) \cdot \mathbf{1} ) $ (eq. 2.) \hspace{1cm} \# Offline NAM prediction.

\State Calculate sample entropy, $H(x_t)$.

\If{$\hat{y} = 1 \textbf{\&} H(\hat{y}_t)<H_0 $}
    \State $n^1_t  \gets n^1_{t-1}  + 1$ ,  $\mu^1_{t} \gets (1 - \frac{1}{n^1_t })  \mu^1_{t-1} + \frac{1}{n^1_t } \frac{f_\theta(x_t)}{||f_\theta(x_t)||}$
\EndIf
\If{$\hat{y} = 0 \textbf{\&} H(\hat{y}_t) < H_0 $}
    \State $n^0_t  \gets n^0_{t-1}  + 1$ , $\mu^0_{t}  \gets (1 - \frac{1}{n^0_t }) \mu^0_{t-1} + \frac{1}{n^0_t }  \frac{f_\theta(x_t)}{||f_\theta(x_t)||}$

\EndIf
\State \textbf{return} $\sigma \left( f_\theta(x_t) \cdot (\mu^1_{t} - \mu^0_{t}) \right)$
\end{algorithmic}
\label{alg:onlinenam}
\end{algorithm}

It is important to note that if the prediction entropy ($H(\hat{y}_t)$) exceeds a certain threshold ($H_0$), the model will skip the update to avoid incorporating inaccurate predictions. A well-trained model typically maintains low prediction entropy. If high entropy persists during test time, it may indicate the need for model retraining or a significant distribution shift between the test and training data.

In addition to \gls{T3A}, we evaluated the effectiveness of EATA \cite{etta}, Tent \cite{tent}, and the update rule from \cite{soul}, referred to as Binary Cross Entropy Threshold (BCT), as alternative methods for updating the weights of the last layer of \gls{nam}. Furthermore, this framework is applicable to both NAM and MT-NAM variants, regardless of whether the $f_\theta$ values are derived from a tree-based model or feature networks.

\begin{table}[!h]
\begin{center}
    \caption{\label{table:table1} Dataset Partitioning: Distribution of sessions and total seizure duration in training, testing, and validation sets per patient (ID: patient number, Sess.: number of sessions, Seiz.: duration of seizures).}
 \vspace{2mm}
\begin{tabular}{ l||cc|cc|cc  }
 \hline
 \multirow{2}{*}{ID} & \multicolumn{2}{c|}{Train}& \multicolumn{2}{c|}{Validation}& \multicolumn{2}{c}{Test} \\
  & Sess.(\#) & Seiz.(s) &  Sess.(\#) & Seiz.(s) & Sess.(\#) & Seiz.(s)\\
  \hline
    1 & 13 & 67 & 2 & 27 & 12 & 375 \\ 
    2 & 16 & 82 & 1 & 82 & 19 & 9 \\ 
    3 & 17 & 238 & 3 & 52 & 4 & 164 \\ 
    5 & 16 & 225 & 1 & 110 & 23 & 333 \\
    6 & 12 & 100 & 2 & 16 & 7 & 53 \\ 
    7 & 11 & 86 & 3 & 86 & 4 & 239 \\ 
    8 & 3 & 171 & 2 & 171 & 15 & 748 \\ 
    9 & 7 & 64 & 2 & 64 & 12 & 212 \\ 
    10 & 20 & 170 & 2 & 65 & 3 & 277 \\ 
    11 & 32 & 22 & 1 & 22 & 1 & 784 \\ 
    12 & 7 & 71 & 1 & 173 & 4 & 202 \\ 
    14 & 11 & 111 & 2 & 41 & 14 & 58 \\ 
    15 & 19 & 608 & 3 & 205 & 17 & 1384 \\ 
    16 & 11 & 18 & 3 & 9 & 3 & 51 \\ 
    17 & 3 & 63 & 1 & 16 & 5 & 170 \\ 
    18 & 30 & 148 & 2 & 68 & 3 & 169 \\ 
    19 & 27 & 78 & 1 & 78 & 1 & 158 \\ 
    20 & 12 & 98 & 1 & 39 & 17 & 196 \\ 
    21 & 18 & 56 & 1 & 56 & 13 & 143 \\ 
    22 & 19 & 58 & 3 & 58 & 9 & 146 \\ 
    23 & 2 & 113 & 1 & 133 & 6 & 291 \\ 
    24 & 8 & 182 & 2 & 19 & 16 & 329 \\
    \hline
\end{tabular}
\end{center}
\end{table}

\begin{figure}[h]
\begin{minipage}[t]{1\columnwidth} 
  \includegraphics[width=\linewidth]{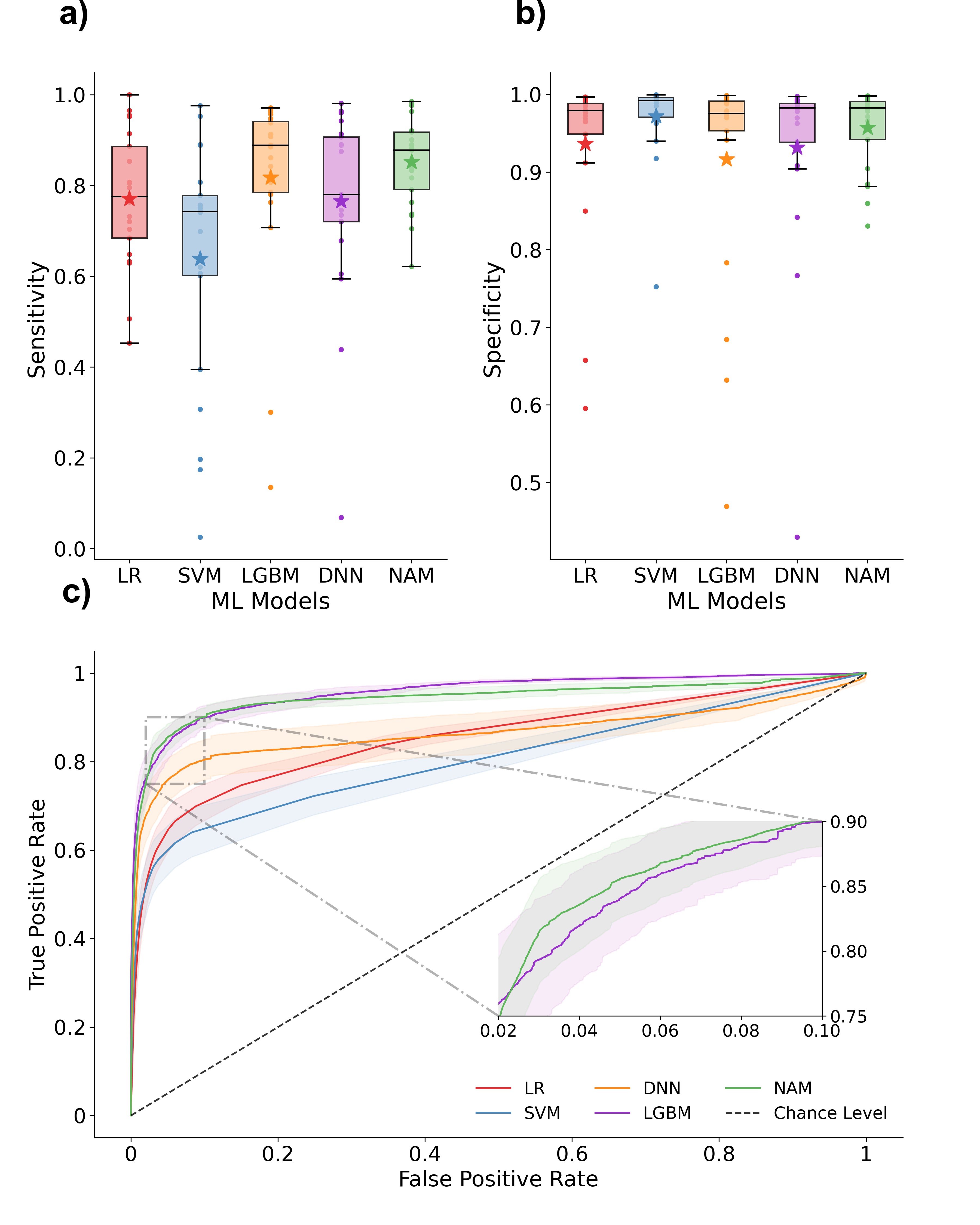}
  \caption{\textbf{(a)-(b)} Window-based sensitivity and specificity across all subjects for the NAM and baseline models. \textbf{(c)} ROC curves (averaged across all subjects) for the NAM and baseline models. The zoomed-in area highlights the performance of NAM and LGBM at low false positive rates.}
  \label{fig:fig2}
\end{minipage}
\end{figure}

\section{Data Pre-Processing, Model Details, and Benchmarks} 
\label{sec:3}
\subsection {Experimental Setup}
\textbf{Dataset.} In this study, we utilized the \gls{chb} scalp EEG dataset \cite{chb-dataset}, which includes recordings from 24 patients, each with 9 to 42 sessions sampled at 256 Hz. The dataset contains a total of 192 seizures. For most patients, we used 23 channels based on the standard 10-20 system. However, for patients with more than 23 channels available, we included all available channels in our analysis. Patient 4 and Patient 13 were excluded from the study due to channel changes during their sessions, which prevented the establishment of a consistent validation set. For Patient 12, only the last session was used, as earlier sessions had shifted EEG montages and only the final session contained sufficient seizure activity for effective model training.

To prevent data leakage during training, we avoid shuffling the data. This ensures that the model does not have access to
seizure samples from future sessions while attempting to predict seizures within the same session \cite{shoaran2018energy,dan2024szcore}. 

\textbf{Data pre-processing.} We extract line length, variance, and spectral band powers from 1-second non-overlapping windows for each patient, as these features have proven effective for seizure detection \cite{OSD, resot, 2022JSSC_Uisub, shin2022256}. Spectral band powers are computed for seven frequency bands: $\delta$ (1-4~Hz), $\theta$~(4-8~Hz), $\alpha$ (8-13 Hz), $\beta$ (13-30 Hz), low-$\gamma$ (30-50 Hz), $\gamma$ (50-80 Hz), and high-$\gamma$ (80-120 Hz). Higher frequencies cannot be extracted due to the dataset’s sampling frequency of 256 Hz, which limits frequency detection based on the Nyquist theorem. These band powers followed by variance and line length will result in 9 features for each channel of EEG for each window of 1 second.

\looseness=-1\textbf{Data split.} In the CHB-MIT dataset, we allocate 15\% of the data for training, 15\% for validation, and 70\% for testing, whenever applicable. Due to the limited number of seizure events in the dataset, we ensure that at least one seizure event is included in both the training and validation sets. Additionally, for each seizure event  in the training set, all preceding non-ictal sessions (sessions without seizures) are also included to maintain continuity in the dataset. To address class imbalance, non-ictal data points are randomly downsampled at a 1:10 ratio compared to ictal data points. This approach helps mitigate the imbalance and enhances the training process. The durations of the training, testing, and validation sets for each patient are detailed in {Table} \ref{table:table1}. Furthermore, since seizure data is time-series data, it is crucial to maintain the chronological order when splitting the dataset into training, validation, and test sets. This ensures that future data points are not used for training the model, while past data points are utilized for validation and testing (see Table \ref{table:table1}).

\begin{figure*}[h]
\begin{minipage}[t]{2\columnwidth} 
\includegraphics[width=\textwidth,height=7cm]{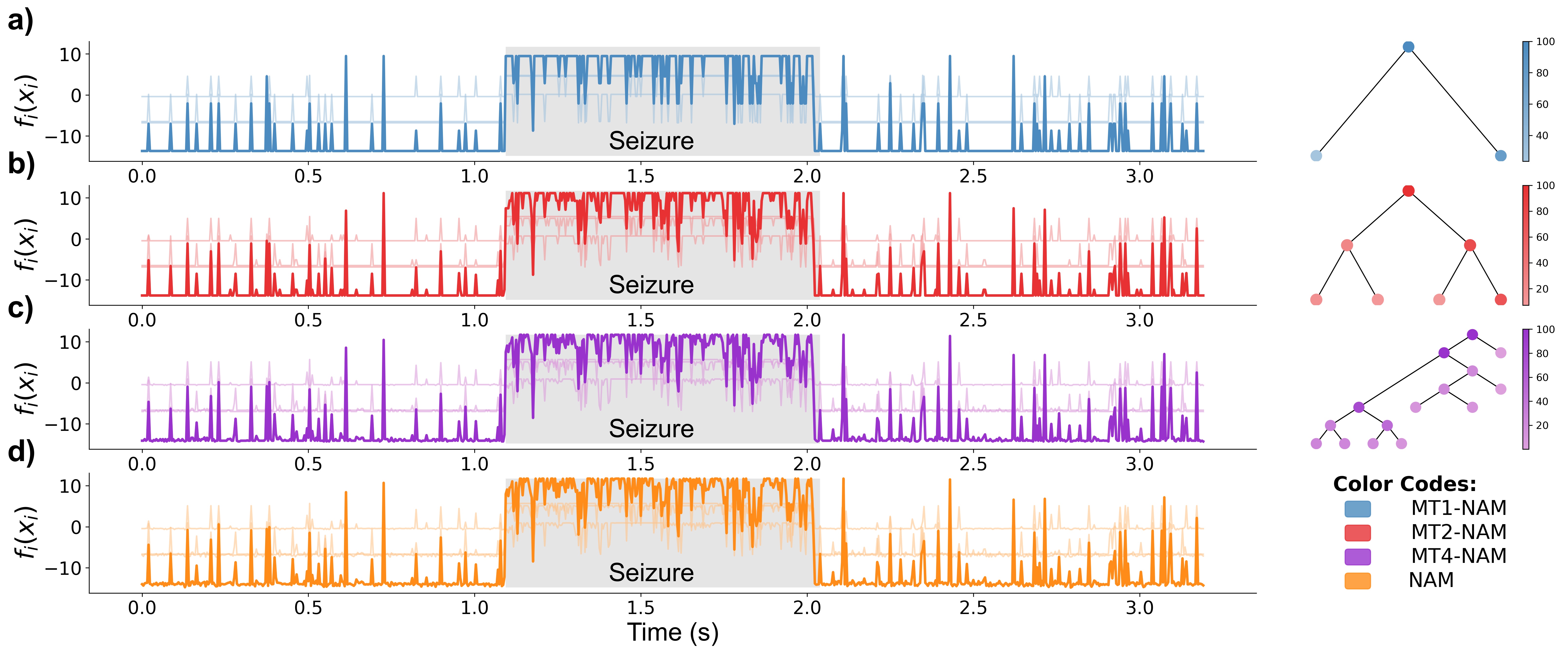}
  \caption{Comparison of the outputs of \textbf{(a)} MT1-NAM, \textbf{(b)} MT2-NAM, \textbf{(c)} MT4-NAM, and \textbf{(d)} NAM. The left plots depict the outputs of three randomly selected feature functions (transparent lines) and their summation (opaque lines). On the right, an example of the micro decision tree used for approximation is displayed, with the color bar representing the proportion of samples passing through each node of the micro tree. } 
  \label{fig:mtnamtrees}
\end{minipage}
\end{figure*}
\begin{figure*}[h]
\begin{minipage}[t]{2\columnwidth} 
  \includegraphics[width=\textwidth]{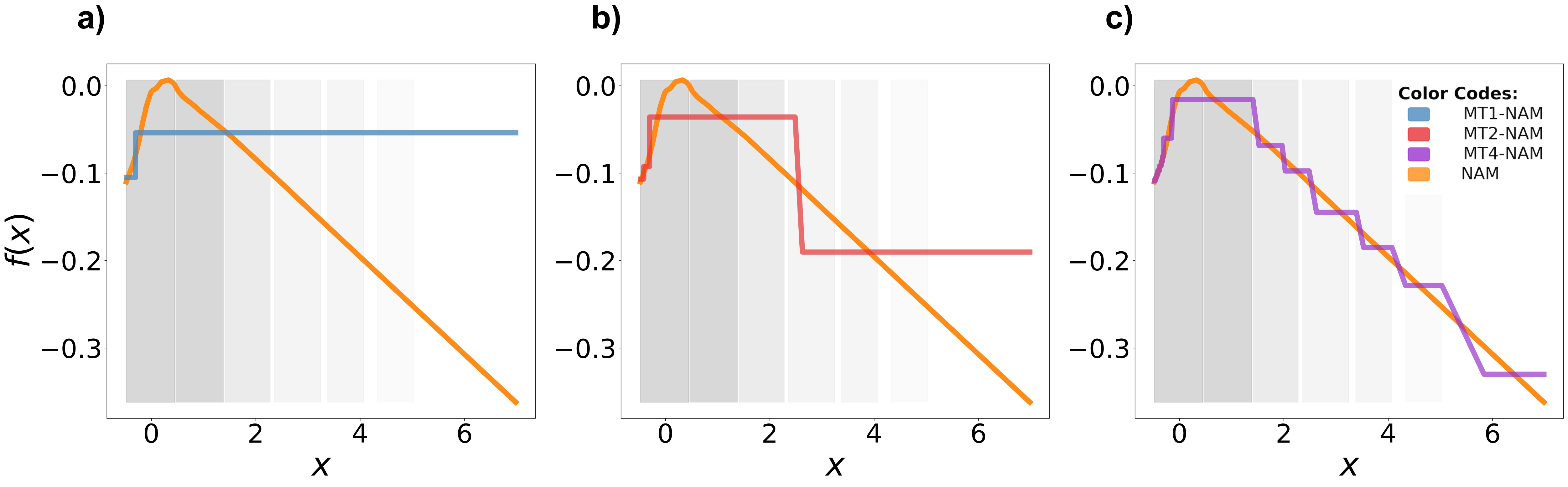}
  \caption{Illustration of a feature function learned by \textcolor{black}{NAM} ($f(x)$) and its approximation ($\tilde{f}(x)$) using \textbf{(a)} \textcolor{black}{MT-1}, \textbf{(b)} \textcolor{black}{MT-2}, and \textbf{(c)} \textcolor{black}{MT-4} approximations of NAM. The white regions in the plots correspond to regions with
low data density (typically a few points) and the gray regions correspond to regions with high data density.} 
  \label{fig:featfunc}
\end{minipage}
\end{figure*}

\textbf{ML model baselines.} To compare the performance of our model with alternative machine learning models, we benchmark NAM against commonly used ML baselines for seizure
detection, including: 
(a) logistic regression (LR) \cite{soul} 
(b) support vector machine (SVM) \cite{svm,circ1} 
(c) deep neural network (DNN) \cite{nnsirc}, and
(d) tree-based models \cite{resot, shoaran2018energy, 2022JSSC_Uisub}.
Given the diversity of tree-based model variations, we  use LightGBM \cite{lgbm}, a state-of-the-art representation of tree-based models.
These baselines are trained and  evaluated on the same pre-processed data, using the same train-test split. All  models are tuned on the same validation set for each patient, with F1-score  used for performance optimization. 
\\
We employed window-based specificity and sensitivity metrics to evaluate the model, measuring the number of correct predictions for ictal and non-ictal windows (using a 1-second window size in this work) across all ictal and non-ictal windows. Additionally, following prior work \cite{soul}, we calculated event-based sensitivity, which measures the proportion of seizure events correctly detected out of all seizure events. To further validate our results, we used AUROC and F1-Score for seizure detection.

{\textbf{Details of Model Training Procedures and Hyperparameters}}
During the model training procedures, we conducted a hyperparameter search on the validation set for both the NAM and baselines. The specific hyperparameters explored for each model are as follows:

\looseness=-1\textit{NAM:}
a) Number of neurons for the first layer of each feature network: 10, 50, 100, 200
b) Activation functions for the first layer: Rectified Linear Unit (ReLU) and Exponential Linear Unit (EXU)
\\
\textit{Logistic Regression:}
Regularization term for the L2 norm: 0.01 to 1
\\
\textit{SVM:}
a) Kernel type: Radial Basis Function (Rbf) and Linear
b) Regularization term for weight regularization using L2 norm: 0.01 to 1
\\
\textit{DNN:}
a) Number of neurons for the first layer: 50, 100, 200, 300
b) Activation function: Leaky ReLU and ReLU
\\
\textit{LGBM:}
a) Number of estimators: 5 to 200
b) Maximum depth of trees: 2, 4, 6, 8, 10
c) Number of leafs: 5, 10, 20

\looseness=-1For the neural networks (DNN and NAM), we used the ADAM \cite{adam} optimizer with parameters $\beta_1=0.9$ and $\beta_2=0.999$. 
These hyperparameters were selected selected based on maximizing the F1-score. All models were trained using a single NVIDIA A100 GPU.

\textbf{MT-NAM structure.} To assess the performance of MT-NAM with varying tree depths, we replace NAM's feature functions with micro decision trees having maximum depths of 1, 2, and 4. By evaluating the model’s performance at different tree depths, we analyze how tree depth affects the overall performance of MT-NAM.

\vspace{5mm}

\textbf{Test-time adaptation baselines.} To enable our model to adapt its predictions based on incoming test-time data, we evaluate the following updating rules for comparison: (a) \gls{T3A} \cite{ttca}, (b) EATA \cite{etta}, (c) Tent \cite{tent}, and (d) BCT \cite{soul}. We perform hyperparameter tuning for all updating methods on the validation set.  The updated models, along with the final update rule parameters from the validation phase, are then evaluated on the test set to assess their performance during the testing phase.

\subsection {Experimental Results}
\textbf{NAM performance.} Consistent with recent studies \cite{soul,asif2020seizurenet}, we evaluate the performance of NAM using sensitivity, specificity, weighted F1-score, and AUROC as the primary metrics for seizure detection. All metrics are computed using a window-based approach, rather than an event-based one, since the event-based sensitivity of NAM is  100\%. As shown in Fig.~\ref{fig:fig2}, NAM demonstrates superior sensitivity compared to all other models. Additionally, NAM outperforms  all models except for SVM in terms of specificity. The low sensitivity of SVM suggests a tendency to overfit to non-ictal inputs. Furthermore, NAM outperforms all other models in terms of AUC-ROC at low False Positive Rates. While LGBM achieves a higher overall AUC-ROC, NAM strikes a better balance in minimizing the False Positive Rate—critical for avoiding frequent false alarms in seizure detection (see zoomed in Fig.~\ref{fig:fig2}(c)). For a detailed comparison of NAM and the baseline models across the specified metrics, please refer to Table \ref{table:table2}.

\begin{table}[!h]
\begin{center}
 \caption{\label{table:table2} Seizure detection results. The mean and standard deviations are calculated across different patients. }
 \vspace{2mm}
\begin{tabular}{ p{1cm}||p{1.4cm}|p{1.4cm}|p{1.5cm}|p{1.4cm}  }
\hline
 & & &  \\
 Model & Sensitivity & Specificity & F1 Score&AUC-ROC \\
 \hline
 \rowcolor{blue!10}
 \textbf{NAM}  &  \textbf{0.85$\pm$0.09}   &   0.96$\pm$.05   & \textbf{0.653$\pm$0.10} &0.94$\pm$0.05 \\
LGBM& 0.82$\pm$0.20  &   0.92$\pm$0.06 & 0.635$\pm$0.13 & \textbf{0.96$\pm$0.04}\\
 LR&  0.77$\pm$0.14    &  0.94$\pm$0.11  & 0.628$\pm$0.12  &0.85$\pm$0.07\\
 SVM&   0.63$\pm$0.25    &   \textbf{0.97$\pm$0.05}   & 0.631$\pm$0.11 & 0.80$\pm$0.12\\
 DNN& 0.77$\pm$0.21  &   0.93$\pm$0.05 &  0.621$\pm$0.11 & 0.86$\pm$0.14\\
 \hline
 \end{tabular}
\end{center}
\end{table}
\textbf{MT-NAM: A streamlined and efficient distillation of NAM} 
Although NAM outperforms other benchmarks in seizure detection (Table \ref{table:table2}), it experiences significant computational latency during inference due to its deep structure (similar to DNNs), requiring a time-consuming forward pass to generate prediction. This limitation makes it unsuitable for real-time applications. The large number of multiplications and nonlinear operations within the feature networks, which are essentially neural networks used for feature function extraction, contribute to an inference delay of 10 ms on modern GPUs which is significantly higher than data stream frequency (200Hz). While neural networks are effective for fitting feature functions during training, they prove sub-optimal for fast inference.


In order to enhance the efficiency of NAM for real-time applications and future hardware implementations, we explore the use of micro-decision trees to approximate NAM's feature functions. We conduct experiments with  tree of 1, 2, and 4, and evaluate their impact on model accuracy and inference time (Table \ref{table:table3}). Remarkably, we observe a nearly {100$\times$} faster during inference while maintaining high performance. The model's sensitivity decreases by less than 2\%, while specificity improves by approximately 1\% with the use of MT4-NAM. 
\\
Fig.~\ref{fig:mtnamtrees} visually illustrates how the outputs of the replaced feature functions in MT-NAM converge toward the actual NAM feature functions as  tree depth increases. Even with a depth of one, MT1-NAM produces results that closely resemble NAM’s feature functions. As the tree depth increases, the outputs of MT2-NAM and MT4-NAM become nearly indistinguishable from NAM’s feature functions, which use neural networks for feature extraction. 
This similarity indicates that the model's performance remains consistent when using MT-NAM instead of NAM. 
Therefore, we conclude that MT4-NAM achieves comparable accuracy while providing a highly efficient model suitable for low-power hardware platforms and real-time applications. Fig \ref{fig:featfunc} illustrates how the original feature function of NAM $f(x)$ is approximated ($\tilde{f}(x)$) by increasing the tree depth which results in more aligned approximation of the function $f(x)$ learned by NAM. this also demonstrates that even decision trees with a depth of 2 can serve as an acceptable approximation (as shown in Table \ref{table:table2}), since the data density in regions where the MT2-NAM (red line) diverges from the original NAM (orange line) is significantly lower compared to other regions. The data density is illustrated by the gray regions in the background of Figure \ref{fig:featfunc}.
\\

\begin{figure*}[h]
\begin{minipage}[t]{2\columnwidth} 
\includegraphics[width=\textwidth,height=7cm]{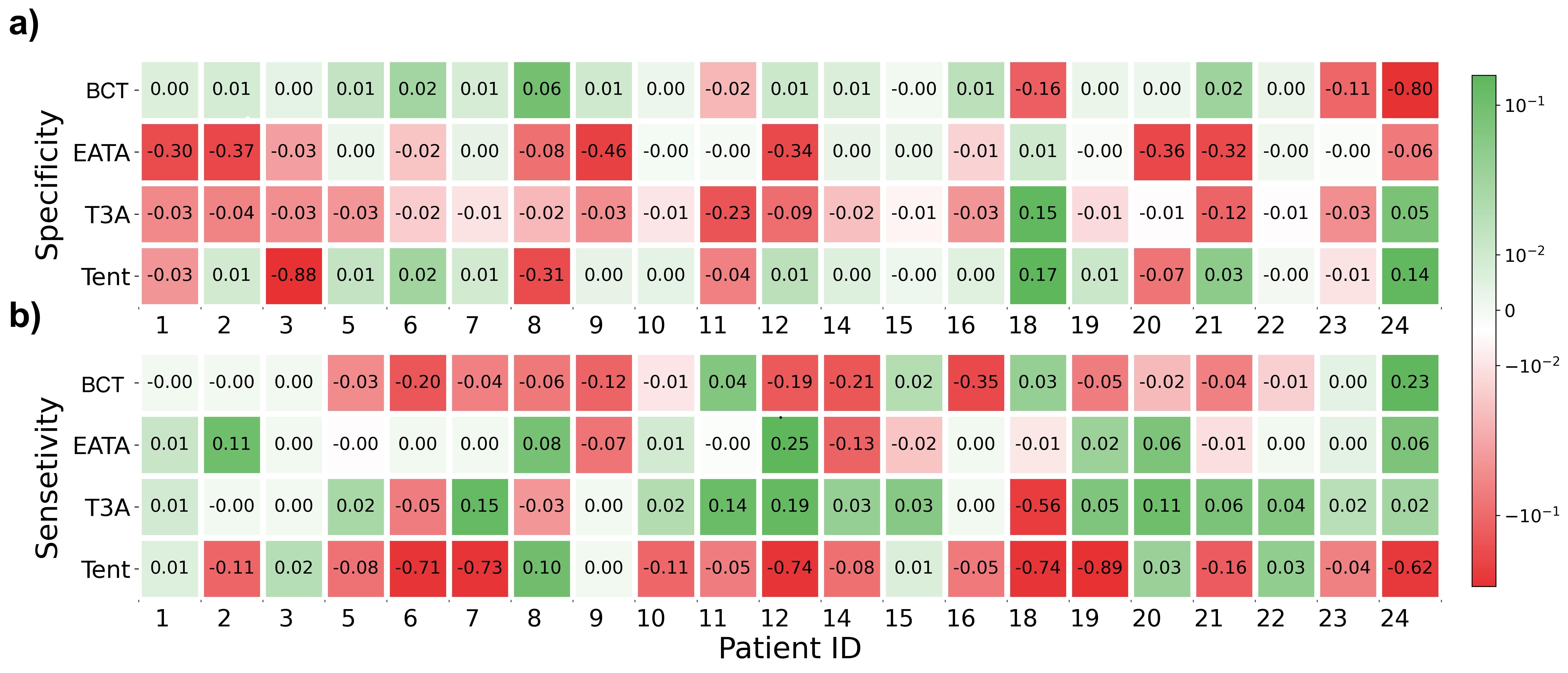}
  \caption{Per-patient analysis of window-based specificity and sensitivity across different online update methods. The results indicate that, among all updating approaches, T3A demonstrates the greatest improvement in seizure sensitivity. The red color indicates the drop and green color indicates the increase in the metric over the offline performance after the online update is applied.} 
  \label{fig:onlineall}
\end{minipage}
\end{figure*}
\begin{figure}[h]
\begin{minipage}[t]{1\columnwidth} 
  \includegraphics[width=\linewidth]{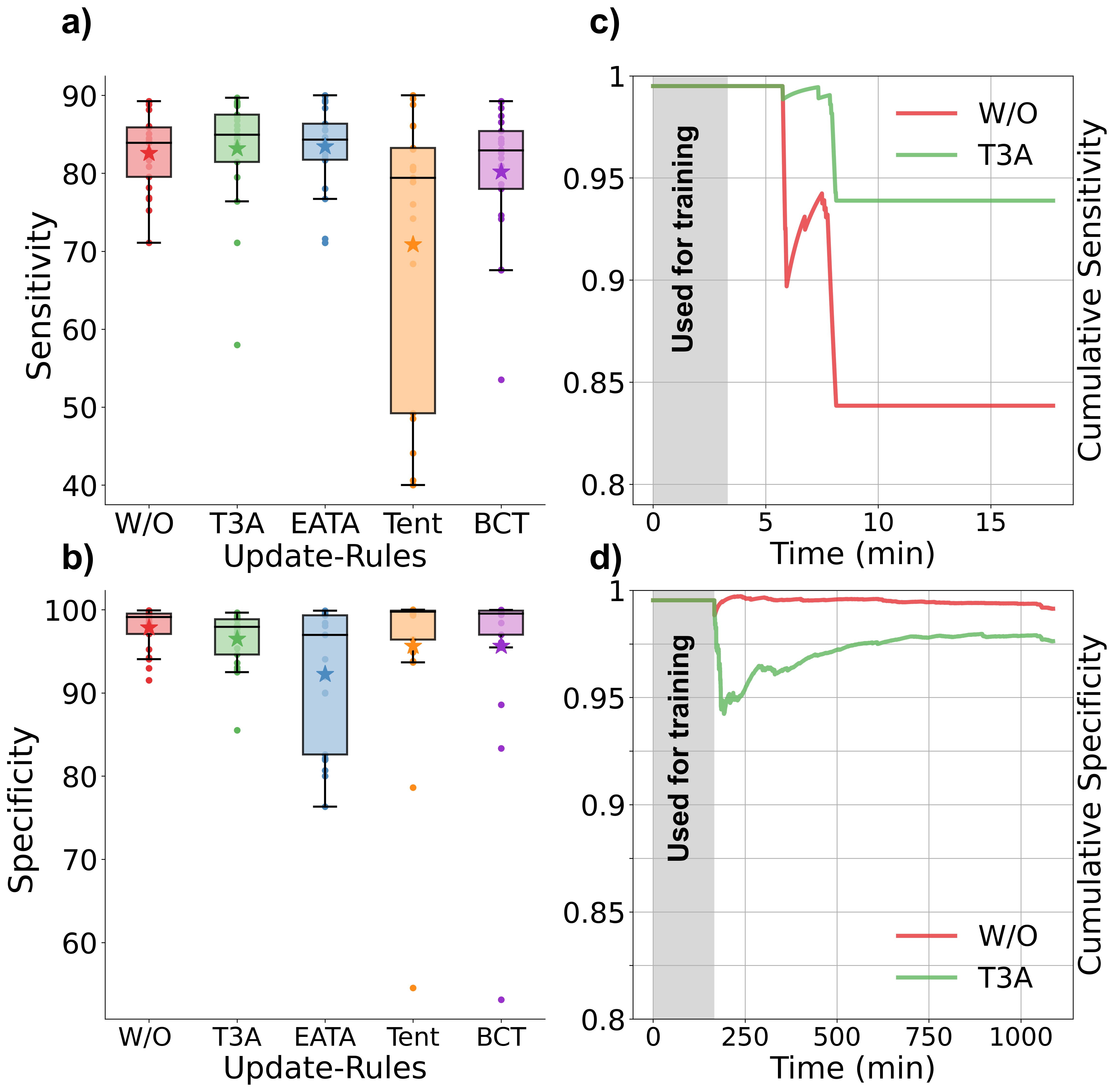}
  \caption{\textbf{(a)} Comparison of sensitivity for NAM, without and with different update rules.
\textbf{(b)} Comparison of specificity for NAM, without and with different update rules.
  \textbf{(c)} Cumulative sensitivity over time for subject \#7, with and without the T3A update.
\textbf{(d)} Cumulative specificity over time for subject \#7, with and without the T3A update. Error bars are representing the standard error and stars are representing the mean value of the metric.
}
  \label{fig:onlinecompare}
\end{minipage}
\end{figure}

\looseness=-1\textbf{Test-time adaptation enhances MT-NAM's performance.} 
As noted in the previous section, the use of MT-NAM feature functions significantly improves inference speed. However, this approximation leads to a 2\% reduction in model sensitivity. Since missing a seizure window can be critical, improving sensitivity is crucial. This challenge can be addressed through various update techniques. Therefore, we explore several test-time adaptation methods in the following section.

We evaluate the performance of four distinct update methods, namely EATA \cite{etta}, Tent \cite{tent}, T3A \cite{ttca}, and BCT \cite{soul}, for adapting our model during test-time. These evaluations are performed on NAM without feature function approximation to identify the most effective approach for updating MT-NAM during inference (Fig.~\ref{fig:onlinecompare}(a), (b)). 

Furthermore, as outlined in \cite{tent}, only the last layer of NAM is updated during the test phase. This approach is chosen to avoid the computational complexity of backpropagating updates through the entire network in real-time applications. Additionally, this strategy aligns with NAM's architecture by focusing on the output of the feature functions, enhancing interpretability and highlighting the importance of specific input features during inference.

\begin{table}[h]
\begin{center}
 \caption{\label{table:table3} Performance and run-time comparison of different NAM Models. Mean and standard deviations are calculated across different patients for sensitivity, specificity and inference time.}
  \vspace{2mm}
\begin{tabular}{l||l|l|l}
 
 \hline 
 \multirow{2}{*}{Model} &  \multirow{2}{*}{Sensitivity} &  \multirow{2}{*}{Specificity} & \multirow{2}{*}{\thead{Inference \\ time (ms) }} \\ 
  & & &  \\
 & & &  \\
 \hline
 NAM  &  {0.85}$\pm${0.09 }   &   $0.96\pm0.050$  &  $10\pm2$ \\
 MT1-NAM   &  $0.70\pm0.22$    &  $\textbf{0.98}\pm\textbf{0.03}$  & \textbf{0.102}$\pm$\textbf{0.02 } \\
 MT2-NAM &   $0.80\pm0.14$    &   $0.97\pm0.03$    &\textbf{$0.107\pm0.02$}  \\
 MT-4 NAM  & $0.83\pm0.12$  &   $0.97\pm0.03$  &\textbf{$0.108\pm0.03$} \\
\hline 
NAM w/ T3A & \textbf{0.86}$\pm$\textbf{0.14}  &  {0.93}$\pm${0.06}    & 10 $\pm$ 2 \\
MT1-NAM w/ T3A &  $0.75\pm0.08$   & $0.96\pm0.03$  & {0.24}$\pm${0.08 }\\
MT2-NAM w/ T3A &   $0.84\pm0.12$  & $0.94\pm0.07$    & {0.24}$\pm${0.08 } \\
 \rowcolor{violet!17}
MT4-NAM w/ T3A & \textbf{0.86}$\pm$\textbf{0.14} & {0.93}$\pm${0.07}   &{0.24}$\pm${0.08} \\
\hline 
\end{tabular}
\end{center}
\end{table}

\begin{figure*}[h]
\begin{minipage}[t]{2\columnwidth} 
  \includegraphics[width=\textwidth]{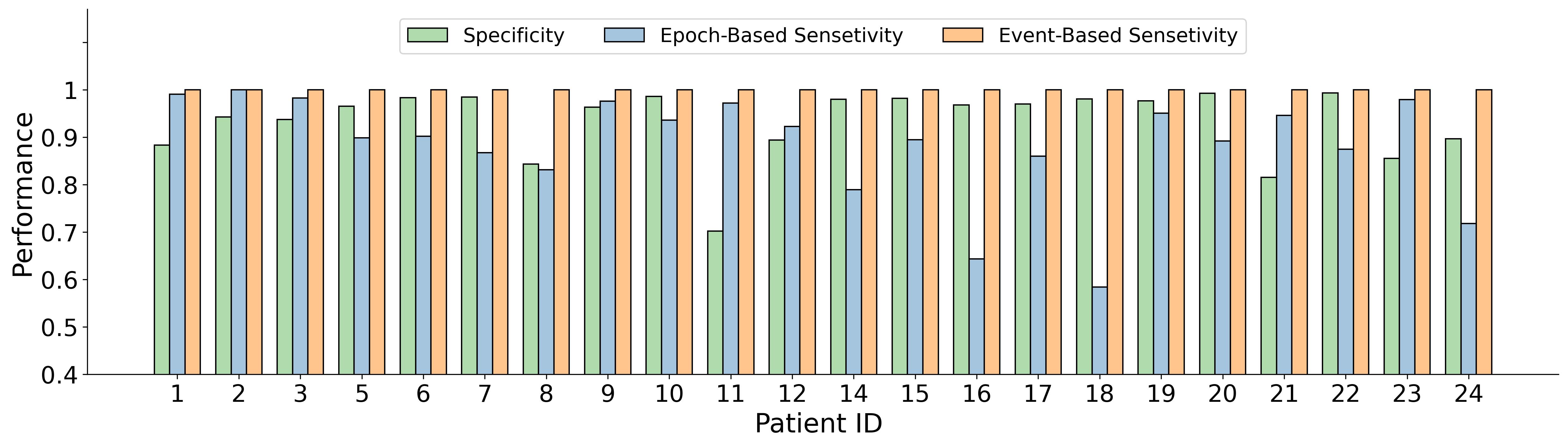}
  \caption{Specificity, window(epoch)-based, and event-based sensitivity across all patients for MT4-NAM with the T3A update rule.} 
  \label{fig:allresbar}
\end{minipage}
\end{figure*}

Among the update rules included in our study, T3A was the only one that improved sensitivity without significantly sacrificing specificity. In Fig.~\ref{fig:onlinecompare}(c) and (d), we observe that with the T3A update method, the cumulative sensitivity for a selected patient increases over time, while the reduction in specificity remains minimal. These results suggest that T3A enables the model to adapt to unseen data during test time by adjusting predictions, improving the model's detection of ictal samples. As shown in Table \ref{table:table3}, extending MT4-NAM with T3A as the test-time update rule achieves accuracy comparable to NAM without feature function approximation, but with a significantly shorter inference time and 50$\times$ faster than original NAM. 
Patient-specific results are presented in Fig.~\ref{fig:onlineall}, illustrating the trade-off between specificity and sensitivity across all updating rules. T3A achieves the highest sensitivity with nearly the same specificity as the offline version of NAM, demonstrating its strengths and stability over other techniques. 
\\
The overall performance analysis for each patient in the CHB-MIT dataset is presented in Figure \ref{fig:allresbar}, which highlights the window-based and event-based sensitivity, as well as window-based specificity, for each patient. Notably, MT4-NAM successfully detects all seizure events, achieving 100\% event-based sensitivity. However, since this metric only evaluates whether a seizure event is correctly detected, we also provide a more detailed analysis through the window-based sensitivity and specificity results of MT4-NAM (Figure \ref{fig:allresbar}), demonstrating its superior performance. These results are also reported as averages in Table \ref{table:table3}.

\newpage

\begin{figure}[h]
  \includegraphics[width=0.5\textwidth]{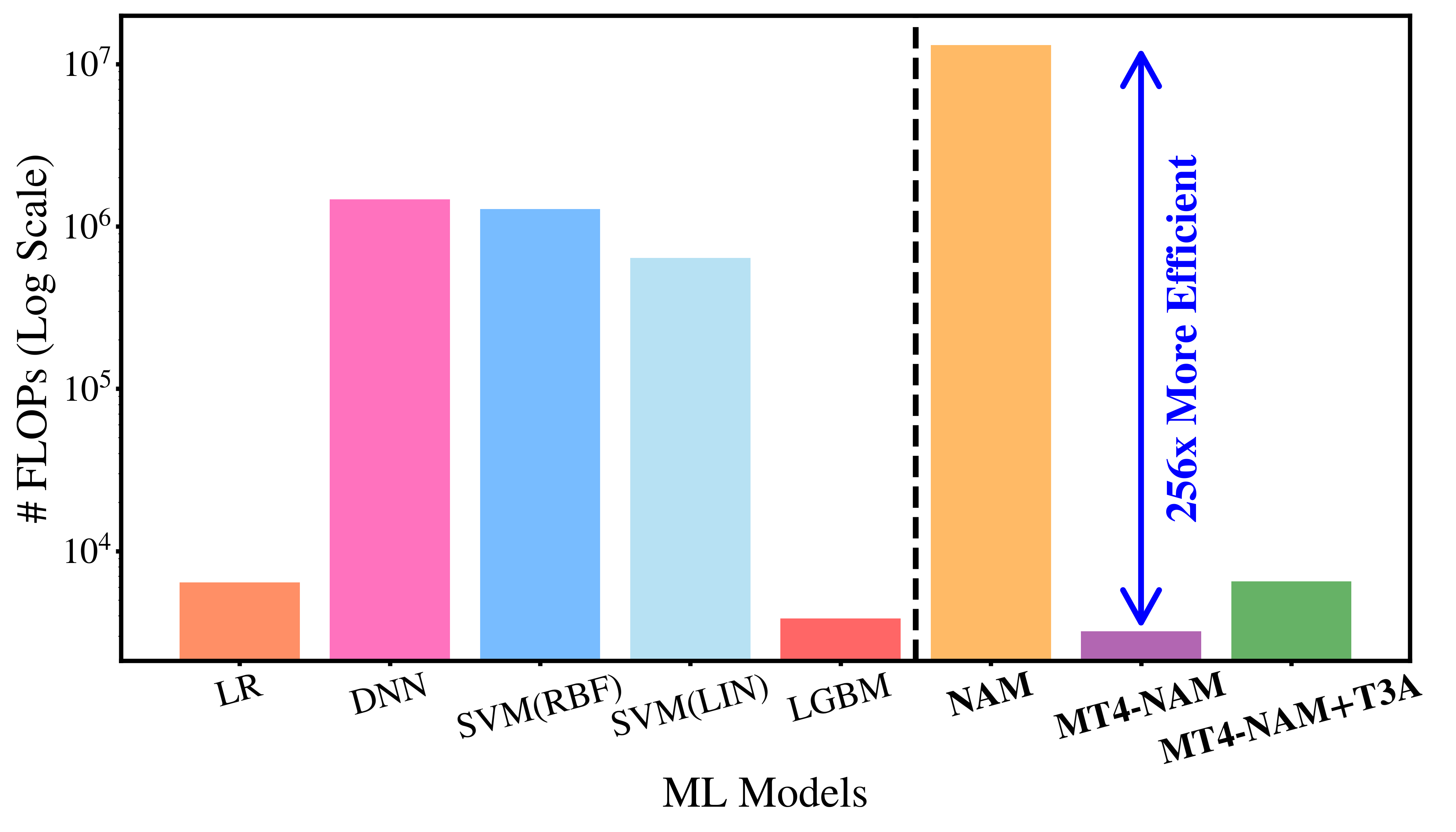}
  \caption{Number of FLOPs for NAM, MT4-NAM, MT4-NAM+T3A and baselines (y-axis in log scale).} 
  \label{fig:efficiy}
\end{figure}

\textbf{Efficiency Comparison}
We compared the efficiency of NAM, MT4-NAM, and MT4-NAM+T3A against baseline models. As shown in Figure \ref{fig:efficiy}, while NAM demonstrates high accuracy, it requires significantly more floating point operations (FLOPs) and computational resources due to its architecture, where each input feature is modeled by an individual feature network. This design leads to quadratic computational complexity with respect to the model size \( d \). To address this challenge, MT4-NAM distills these feature networks into small regression trees, eliminating the need for complex computations in the regression network and relying instead on simple comparisons, similar to other tree-based models. This \textit{tree-based distillation} approach results in a 256-fold improvement in computational efficiency, making the model far less demanding in terms of computation.

As shown in Figure \ref{fig:efficiy}, MT4-NAM exhibits the lowest computational overhead among all models tested. Moreover, MT4-NAM with the T3A update mechanism achieves computational complexity comparable to logistic regression while offering significantly higher accuracy. This demonstrates that MT4-NAM+T3A is a promising solution for efficient and accurate seizure detection and underscores the potential of \textit{tree-based distillation} as a powerful method for knowledge distillation in neural additive models.

\section{Related Work}
\label{sec:relwork}
This section reviews previous work on models for seizure detection, including approaches in knowledge distillation and online updating methods used in this domain:
\\
\\
\textbf{Seizure Detection Models:} Various models have been proposed for seizure detection. For example, \cite{cnnlstm} applied a CNN-LSTM approach for patient-independent seizure detection, while \cite{rnn1} explored RNN-based methods. Additionally, \cite{svmmodel, svm} employed SVM-based learning algorithms, \cite{cnn1} utilized a CNN model, and \cite{li2022graph} proposed a deep graph neural network for improved detection. 
\\
\\
\textbf{Online Learning Methods for Epilepsy:} While much of our focus has been on test-time adaptation methods commonly used in software literature, such as \cite{ttt, tent}, we also emphasize efforts in hardware-based algorithms for seizure detection, as explored in \cite{soul, closedloop}, which serve as key comparisons in our study. Studies like \cite{zshot} apply zero-shot fine-tuning for each patient to enhance performance at test time, \cite{adaptive} utilizes adaptive learning techniques for seizure detection, and \cite{cluster} employs a CNN-based model with hybrid feature selection and one-shot learning to improve detection accuracy. \cite{amirhoss} used continual learning for patient specific seizure detection.
\\
\\
\textbf{Knowledge Distillation: }As discussed in Section \ref{sec:2}, studies applying knowledge distillation to seizure detection primarily use teacher-student models, where the student model is retrained with labels provided by the teacher. For instance, \cite{multi} implemented a multi-teacher to single-student approach, and \cite{gnndist} applied distillation techniques specifically for deep graph neural networks to improve seizure detection.

\section {Conclusions}
\label{conc}
Our study introduces a novel approach to EEG-based seizure detection using the Neural Additive Model (NAM) architecture. The proposed method outperforms commonly used hardware-implemented baselines for seizure detection. Leveraging the interpretability of the NAM architecture, we introduce micro decision trees as an efficient, distilled alternative to the heavily parameterized feature networks of NAM. Experimental results demonstrate the effectiveness of MT-NAM in seizure detection, achieving 50 $\times$ higher speed during inference time without sacrificing performance. By incorporating the T3A update rule with the MT-NAM structure, we effectively utilize test-time data to enhance the model's sensitivity to seizure events, while minimizing the impact on specificity. 
These findings highlight the potential of MT-NAM combined with the T3A method for real-time applications and hardware implementations, offering greater adaptability for next-generation brain stimulation devices.

\section{Acknowledgments}

This work was supported in part by the Swiss State Secretariat for Education, Research and Innovation under Contract number SCR0548363, and in part by the Wyss project under contract number 532932 and the Swiss National Science Foundation (SNSF) under grant number 200021-205011.

\bibliographystyle{IEEEtran}
\bibliography{refs}

\end{document}